\documentclass{aa}
\usepackage{graphicx}

\begin{document}

\title{Self-similar condensation of  rotating magnetized
self-gravitating isothermal filaments}

\titlerunning{Self-similar condensations of  self-gravitating filaments}

%\authorrunning{P. Hennebelle}

\author{P. Hennebelle}

\offprints{P. Hennebelle, \email{patrick.hennebelle@ens.fr}}

\institute{ Laboratoire de radioastronomie millim{\'e}trique, UMR 8540 du
CNRS, {\'E}cole normale sup{\'e}rieure et Observatoire de Paris,
 24 rue Lhomond, 75231 Paris cedex 05,
France \and Department of Physics and Astronomy, Cardiff University, PO Box 913,
 5 The Parade, Cardiff CF24 3YB, Wales, UK
}

\abstract{Ordinary differential equations 
describing the self-similar collapse of a 
rotating, magnetized, self-gravitating and isothermal filament
are derived. Explicit homologous solutions are studied 
with special emphasis on the bifurcation that occurs at the
magnetosonic critical point. It is shown that there is a critical value
for the toroidal magnetic field  slope at the origin  above which no 
bifurcation occurs, the solution remains homologous,  and below which
the density and the poloidal magnetic field tend to zero
 at large radius (envelope) whereas the toroidal magnetic field and azimuthal 
velocity relax towards a constant value. 
A series of spatial profiles of density, velocity and magnetic field,
 potentially useful for comparison with numerical
or observational studies, is  obtained numerically and discussed.
%When the toroidal field is weak, the density field in the envelope
% is weakly peaked whereas
%when it is near the critical value, the density profile is  stiffer and the
%azimuthal velocity and magnetic toroidal field profiles flatter.
}

\maketitle

\section{Introduction}

Understanding the physics of dense cores is a crucial issue for star 
formation. In this respect the shape of the cores is of great interest 
since it depends on the physics that drives their evolution.

%The standard theory   of low mass star formation based on ambipolar 
%diffusion, predicts that the dense cores should be oblate and that the 
%magnetic field should be parallel to the minor axis of the cores.
%However, 

Based on a statistical shape analysis,
 various studies have shown that the observational
 axis ratio distributions
derived from a catalog of dense cores (e.g. Jijina et al. 1999)
are not compatible with a distribution of oblate cores and compatible or 
marginally
compatible with a distribution of prolate cores 
 (David \& Verschueren 1987, Myers 1991, Ryden 1996, Curry 2002). 
The same conclusion is reached by Hartmann (2002) who shows that 
the major axis of the cores in Taurus is preferentially aligned with the 
filaments in which they are embedded.

%Recent polarization measurements by Ward-Thompson et al (2000) 
%(see also Lai et al. 2001) find an 
%angle between the projected minor axis and the field direction of 
%about $30^\circ$ which is also 
%a departure from  the standard theory of low mass stars.

 Interestingly Jones et al. (2001)
 and Goodwin et al. (2002) have 
shown recently that the apparent axis ratio distribution of the dense cores is
compatible with the cores being triaxial but being more oblate than
prolate. However
the physical origin of such structures  is presently unclear and the prolate 
assumption appears thus to be the simplest one which is compatible with
the current observations.

%%Basu (2000) also argue that the
%offset angle between the major axis and the polarization direction
%observed by Ward-Thompson et al. (2000) in various dense cores
%can be explained as well 
%by the fact that the dense cores are triaxial (see also Galli et al. 2001).
%This suggests that only  minor deviations of the standard theory could 
%explained these observations.

%On the other hand an alternative explanation is that the dense cores are 
% prolate structures that maybe  permeated by an helical magnetic field.

 It is the purpose
of the present study to investigate further the collapse of such elongated
structures.

\subsection{Prolate dense cores}
Static prolate cores have been found in various theoretical studies. 

Curry (2000) has numerically obtained  solutions of the Lane-Emden
equations having a prolate shape. These solutions are periodic 
along the z-axis and reminiscent of the quasi periodicity of the cores
sometimes observed in filaments (e.g. Dutrey et al. 1991).

Curry \& Stahler (2001) have shown that  prolate cores 
can be  solutions of the static MHD equations if the structure
of the poloidal magnetic field is such that it compresses the cloud.

Finally cores permeated by an helical magnetic field can also be prolate 
 (Tomisaka 1991, Habe et al. 1991, Fiege \& Pudritz 2000c).
Fiege \& Pudritz (2000c) also show that a large sample of  cores 
permeated by an helical magnetic field 
 have a shape compatible with the observational axis ratio distribution
and that the polarization hole effect (Fiege \& Pudritz 2000d), 
observed towards the densest part of the clouds, 
can be nicely interpreted  as a  
purely geometrical effect due to cancellation of the contribution 
of the poloidal flux in the core and the toroidal flux in the envelope.

On the other hand, prolate cores  can  be the result of an instability 
in magnetized filaments as  first investigated
by Chandrasekar \& Fermi (1953) and  recently by 
  Nakamura et al. (1993) who show 
that a filamentary molecular clouds permeated by an helical field
of various intensity can be instable   through a combination
of the gravitational and the magnetized sausage instability
 leading to the formation of  prolate cores. More recently, Fiege \& Pudritz
(2000a,b) extend this approach to molecular clouds confined by an external 
pressure. 

\subsection{Aim of the paper}
In  this paper we derive self-similar solutions that can describe
the collapse of a magnetized, rotating, isothermal self-gravitating 
filament. We obtain four ordinary equations in  
 $X$, the self-similar variable. These equations are very similar
to the equations  obtained by Larson and Penston for the 
collapse of a spherical non rotating and non magnetized cloud 
(Larson 1969, Penston 1969) and extensively studied by Hunter (1977), 
Shu (1977), Bouquet et al. (1985) and  Whitworth \& Summers (1985).
Our equations admit as exact solutions, some of the explicit
solutions which are obtained by Aburihan et al. (2001) and by the author
(Hennebelle 2001). These solutions present some important 
restrictions since  they are unbounded and the magnetic poloidal pressure 
 vanishes. The present study  demonstrates that some of these
solutions are indeed bounded, since a bifurcation arise at the magnetosonic
point. Semi-analytical solutions that  have a 
non-vanishing magnetic poloidal pressure are also obtained.

Our goal  is to understand better the gravitational contraction
of  a  filament since this process could be relevant in 
the context of  prestellar prolate core collapse, and to 
  obtain the spatial profiles
of the fields resulting from the dynamical collapse of a 
magnetized filament. These profiles could then be compared with 
observational data and guide a full numerical simulation, although 
the filamentary geometry used in the present study is a significant departure
from a prolate configuration.

The second section of the paper presents the formalism and discusses the 
physical meaning of the equations. 
In the third section, we discuss  explicit solutions of these equations 
 and we carefully
address the question of the magnetosonic critical points. 
In the fourth section, we solve numerically the ordinary differential
 equations and discuss the physical meaning of the solutions.
The fifth section concludes the paper.

\section{Self-similar solutions}
\subsection{Reduction to ordinary differential equations}
We consider the perfect MHD equations of a self-gravitating gas in cylindrical
coordinates. We thus ignore the ambipolar diffusion process.

%These equations are  the same that the equations considered 
%  in a previous paper (Hennebelle 2002) and are not  written explicitly 
%here. 

We make the usual self-similar reduction and set, in the usual notations:
\begin{eqnarray}
\nonumber
X &=& { \varpi \over C_s t}, \\
\nonumber
\rho(r,t) &=& {1 \over 4 \pi G} { R(X) \over t^2} , \\
\nonumber
\Phi(r,t) &=& C _s ^2 \phi(X) , \\
\label{champ}
V _\varpi (r,t) &=& C _s v _\varpi (X), \\
\nonumber
V _\theta (r,t) &=& C _s v _\theta(X), \\
\nonumber
V _z (r,t) &=& \alpha  {z \over t} , \\
\nonumber
B _\theta (r,t) &=& \sqrt{\mu _0 C_s^2 / (4 \pi G)} { b _\theta(X) \over t }, \\
\nonumber
B _z (r,t) &=& \sqrt{\mu _0 C_s^2 / (4 \pi G)} { b _z(X) \over t },
\end{eqnarray}
where $\alpha$ is a real number. The radial magnetic component has to 
vanish in order to ensure zero magnetic divergence. 

The problem considered in the present study is thus
axisymmetric and  independent of z
 but for the z-component of the velocity field.
Also our analysis leaves aside the instability 
(e.g. gravitational, kink or sausage instability)
that could develop and lead to a more complex evolution. 

If $t < 0$, these fields describe a contraction whereas they describe a 
 rarefaction if $t > 0$.

In the following, we will consider only the case $t < 0$, i.e. the
core contraction until the singularity formation, since during the
subsequent  evolution, it is expected that a disk will form, as it is
the case in the numerical study of Nakamura et al. (1999), that cannot
be described by the field stated by Eq.~(\ref{champ}).

$R, \phi, v _\varpi, v _\theta, b _\theta$ and $b _z$ are dimensionless 
variables and are  expected to be of order unity. 
The   sound speed, $C _s$,  is  about 0.2 km s$^{-1}$ for a 
dense core having a temperature of 10 K.
 Although in principle the self-similar solutions have no characteristic
scales, realistic initial and boundary conditions limit 
the domain of validity.  
The characteristic spatial scale
is given by $r \leq 0.01-0.1$ pc, the characteristic time scale 
 by $t  \leq 10^{3-5}$ years and the density is 
$> 10^{5-6}$ cm$^{-3}$. 
The typical  magnetic intensity
is about $\simeq 10 \mu$G for a density of $\simeq 10^3$ cm$^{-3}$.
Then for higher densities, the relation~:
\begin{eqnarray}
{B \over B _0} = \left( {\rho \over \rho _0} \right) ^k
\end{eqnarray}
where $1/3 \le k \le 1/2$ is theoretically (Mouschovias 1976, Scott \&
Black 1980) and observationally inferred
(Crutcher 1999). 

$B _z$ is the poloidal magnetic field. It tends usually  to support
the cloud against the gravitational collapse.  
$B _\theta$ is the toroidal magnetic component.
It has not been taken into account in most of the 
 collapse studies of a magnetized dense core. Such fields
can be produced through the stretching of the poloidal magnetic field.
The most important effect of the toroidal component is that it usually
compresses  the gas radially. It can also be a support in the axial direction.

The MHD equations lead with the definitions stated by Eqs.~(\ref{champ})
to the following equations.

Continuity:
\begin{eqnarray}
-2 R - X R' + {1 \over X} \partial _X (X v _{\varpi} R) + \alpha R = 0.
\label{eq1}
\end{eqnarray}

Radial momentum conservation:
\begin{eqnarray}
\nonumber
- X v _\varpi ' + v _\varpi v _\varpi ' - {v _\theta ^2 \over X } &=& \\
 -  {R' \over R} + \phi '&-& {b _\theta \over  X R} \partial _X (X b _\theta)
- { b _z b_z' \over R }.
\label{eq2}
\end{eqnarray}

Azimuthal momentum conservation:
\begin{eqnarray}
- X v _\theta ' + v _\varpi v _\theta ' + {v _\varpi v _\theta \over X} =0. 
\label{eq3}
\end{eqnarray}

Axial momentum conservation:
\begin{eqnarray}
- \alpha  + \alpha ^2 = 0.
\label{eq4}
\end{eqnarray}

Poisson equation:
\begin{eqnarray}
{1 \over X} \partial _X (X \phi'  ) = - R.
\label{eq5}
\end{eqnarray}

Azimuthal induction equation:
\begin{eqnarray}
- b _\theta - X b _\theta ' + \partial _X (v _\varpi b _\theta) + \alpha b _\theta =0.
\label{eq6}
\end{eqnarray}

Axial induction equation:
\begin{eqnarray}
\label{eq7}
- b _z - X b _z ' + {1 \over X} \partial _X (X v _\varpi b _z) =0.
\end{eqnarray}

Eq.~(\ref{eq4}) admits the two solutions: $\alpha=0$ and $\alpha=1$, 
which correspond respectively to a  filament static along the z-axis
 and to a filament collapsing homologously along this axis. 
The first configuration has been investigated by Inutsuka \& Miyama (1992) in
the hydrodynamical case.
 In the magnetized case, a development around $X \simeq 0$ 
(see Eqs.~\ref{hypo}) implies that $b _z$, $v _\theta$ and $b _ \theta$
 cannot be all together different from zero which is an important restriction.
Moreover, it is unlikely that a cloud will collapse in the radial
direction only and not along the z-axis.  
Thus in the following,
we will consider only the second case $\alpha=1$.

With $\alpha=1$, it is easily seen with Eqs.~(\ref{eq1}) and (\ref{eq5}) that,
\begin{eqnarray}
\phi ' =  R (v _\varpi -X).
\label{pot_red}
\end{eqnarray}
It is also easy to see with Eqs.~(\ref{eq1}) and~(\ref{eq7}), that the {\it density}, $R$, 
and the axial {\it magnetic fields}, $b _z$, 
are proportional: $b _z = \Gamma _z R$, where
$\Gamma _z$  is a real number.

With these two last equations, the system of Eqs.~(\ref{eq1})-(\ref{eq7}), reduces
after some algebra to:
\begin{eqnarray}
\nonumber
v _\varpi ' &=& { v _\varpi - X \over 
(v _\varpi - X)^2 -1 - b _\theta^2/R - \Gamma _z ^2 R  } \times \\
 & & \left(  {v _\theta ^2 \over X}  + { 1 \over X}    + R (v _\varpi -X) 
  + { \Gamma _z ^2 R  \over X} - {b _\theta ^2 \over X R}   \right),
\label{eqfin1}
\end{eqnarray}

\begin{eqnarray}
R' = {- v _\varpi ' R \over v _\varpi -X} - {R \over X},
\label{eqfin2}
\end{eqnarray}

\begin{eqnarray}
v _\theta ' = {- v _\varpi  v _\theta \over X (v _\varpi -X) }, 
\label{eqfin3}
\end{eqnarray}

\begin{eqnarray}
b _\theta ' = {- v _\varpi ' b _\theta \over v _\varpi -X}. 
\label{eqfin4}
\end{eqnarray}

\subsection{Asymptotic behavior}
In this section we derive the asymptotic behavior of 
Eqs.~(\ref{eqfin1})-(\ref{eqfin4}) in the limit $X \rightarrow -\infty$.
The asymptotic behavior near the origin $X \rightarrow 0$ will be
 presented and discussed in some details in Sect.~\ref{homolo}.

From Eqs.~(\ref{eqfin1})-(\ref{eqfin4}), it is seen that in the limit 
$X \rightarrow - \infty$, we have:
\begin{eqnarray}
\nonumber
R(X) &\simeq& A_1 /  X , \\
\label{asymptote}
v _\varpi (X) &\simeq&  (A _1 + (b _\theta ^\infty) ^2/ A _1) \ln (|X|) , \\
\nonumber
v _\theta (X) &\simeq& v _\theta ^ \infty 
\exp \left( (A _1 + (b _\theta ^\infty) ^2 / A_1)
 \int ^{|X|}  \ln( u ) / u^2 du \right) , \\
\nonumber
b _\theta (X) &\simeq& b _\theta ^ \infty 
\exp( - (A _1 + (b _\theta ^\infty) ^2 / A _1) / X).
\end{eqnarray}
where $A _1$ is a negative real number.

Consequently at large distance, the azimuthal velocity and 
the toroidal magnetic
component tend to constant values, whereas density (and poloidal magnetic
field) decreases as $1/X$. The radial velocity tends  slowly
 to infinity. This is a consequence of the 
infinitely long filamentary geometry.
Let us recall that the velocity of the Larson-Penston solutions  
tends to a finite value at infinity. 
In case of strong toroidal fields, $v _\theta$ and $b_\theta$ converge
quickly to their asymptotic values. 

In the limit $X \rightarrow -\infty$, we have: \\ $b _z/b _\theta \simeq
\Gamma _z  A_1 b _\theta ^\infty / X \rightarrow 0$ and the dominant forces are
the gravitational force and the toroidal magnetic one.

\subsection{Physical discussions}

\subsubsection{Physical interpretations}
\label{interp}
The system of Eqs.~(\ref{eqfin1})-(\ref{eqfin4}) has a clear physical meaning.
It is very similar to the equations derived by Larson and Penston
 (Larson 1969, Penston 1969). 
 As for the
Larson-Penston equations, thermal pressure
 (second term of the right-hand side of Eq.~\ref{eqfin1})
 and  gravity (third term)
 are included. The solutions also include rotation 
(first term), magnetic poloidal forces
(fourth term) and magnetic toroidal force (fifth term).

The system of Eqs.~(\ref{eqfin1})-(\ref{eqfin4})
 describes the contraction (or expansion)
of a filament with an homologous velocity along the z-axis.
The  gravitational force is zero along 
the z-axis which is consistent with an infinitely long filament. 

In a more
realistic situation, the cloud geometry and thus the the potential
 should evolve with time 
 as  for the solutions obtained by Lin et al.
 (1965) for a  cold and unmagnetized cloud.

The velocity along the z-axis does not depend on $\varpi$, and consequently
the solutions will have no clear physical meaning in the limit 
$\varpi \rightarrow \infty$.
 Asymptotic behaviors in this limit, 
will then not be considered in the following.
Because of the homologous velocity field along the z-axis, the
validity of the solutions is also restricted to a limited domain not 
too far from the $z=0$ plan, e.g. $|z| \le $ few $|X _c|$ ($X_c$ is the 
position of the critical point, see next sections).

The magnetic poloidal field is proportional to density which simply means
that ratio between the mass per unit length ($m=2\pi \int \rho r dr$)
 and the poloidal magnetic flux  ($\psi= \int B _z r dr$)
 is constant into the cloud.
It is worth noting that $m / \psi \rightarrow \infty$ 
when $t \rightarrow 0$. This is due to the fact that part
of the gas flows along the field line.

\subsubsection{Magnetosonic singularities}
Eqs.~(\ref{eqfin1})-(\ref{eqfin4}) become singular if: 
$(v _\varpi - X)^2 -1 - b _\theta^2/R - \Gamma _z ^2 R  \rightarrow 0$.
This is the critical point already obtained in the
Larson-Penston equations. The first term is the square of the fluid velocity
relatively to the similarity profile, the second one is the sound speed, 
the two last terms represent the Alfv\'en velocity, $C _a$.
 The physical meaning
is  that when the fluid velocity relatively to the similarity profile
reaches the fastest velocity at which waves can travel, a singularity
occurs. It is only the solutions that pass this singular point  that 
have a physical meaning.
The difference from the equations obtained by Larson and Penston is that
 since the magnetic field is included, the fastest waves  are the fast 
MHD waves and travel with a velocity equal to: $\sqrt{ C _s ^2 + C _a ^2}$.

In their study, Bouquet et al. (1985) show that for any 
value of $\gamma$, the critical point reached by  the homologous
solutions is a node rather than a saddle. In the isothermal case, 
Whitworth \& Summers (1995) 
(see also Bouquet 1984) were able to carry out a general study of 
the critical point (not restricted to the homologous solution) and show
that if $|X _c| < 1$, the critical point  is a saddle and the solution cannot 
pass it, whereas in the other case, the critical point is a node and
the solutions are physical.

\section{Explicit homologous solutions and magnetosonic critical point}
\label{homolo}
\subsection{Explicit solutions: the homologous sub-Alfv\'enic core}
In this section, we derive explicit solutions of
 Eqs.~(\ref{eqfin1}),~(\ref{eqfin2}),~(\ref{eqfin3}),~(\ref{eqfin4}). These
solutions have already been obtained by the author (Hennebelle 2001)
 in an extended
formulation of self-similarity (Eq. (63) with $\Gamma _z =0$). 
For completeness and clarity we rederive them here.

We look for homologous solutions and  assume the following spatial
dependence: 
\begin{eqnarray}
\nonumber
R(X) &=& R _0, \\
\nonumber
v _\varpi(X) &=& v_{\varpi, 0}' X, \\
\label{hypo}
v _\theta(X) &=& v _{\theta, 0}' X, \\
\nonumber
b _z (X) &=& b _{z,0} = \Gamma _z  R _0  , \\
\nonumber
b _\theta (X) &=& b _{\theta,0}'  X = \Gamma _\phi R _0 X.
\end{eqnarray}
where, as in Fiege \& Pudritz (2000a), 
 $\Gamma _z$ and $\Gamma _\phi$ are respectively the   poloidal 
and  toroidal  magnetic flux to mass ratio.

It is straightforward to show that Eqs.~(\ref{eqfin2})-(\ref{eqfin4}) require
that:
\begin{eqnarray}
\label{cond}
v _{\varpi,0} ' &=& 1/2, 
\end{eqnarray}
whereas Eq.~(\ref{eqfin1}) leads to:
\begin{eqnarray}
1/4 + (v _{\theta, 0}') ^2 &=& R _0 /2 + 2 (b _{\theta,0}') ^2 / R _0. 
\label{sol}
\end{eqnarray}

In Eq.~(\ref{eqfin1}) the terms independent of $X$ cancel out and lead
to no new constraints, whereas the terms proportional to $X ^2$ lead
to the condition stated by Eq.~(\ref{sol}).
Consequently, 
the fields stated by Eqs.~(\ref{hypo}) with the constraint stated by 
Eq.~(\ref{cond}) constitute the asymptotic form of the solutions of 
Eqs.~(\ref{eqfin1})-(\ref{eqfin4}) at $X \rightarrow 0$. 
The homologous solutions  require two constraints 
(Eqs.~\ref{cond} and~\ref{sol}) and there are five parameters, 
which leaves three independent degrees of freedom (e.g. the poloidal and 
toroidal magnetic flux to mass ratio and the density).

 Eq.~(\ref{sol}) has a clear physical meaning. 
It is simply the radial momentum conservation in the pressureless limit.
 The first
term of the left-hand side is the square of $v _{\varpi,0}'$, and represents
 the advection term, the
second one is the centrifugal force, the first term of the right-hand side 
is the gravitational force whereas the second one is the toroidal
magnetic pinching.
 At this point
there is no explicit poloidal support since the axial magnetic component
 is uniform. However, there is an implicit effect due to the
critical point (see Sect.~\ref{crit_point} and~\ref{blabla_crit}). 

If $b _{\theta,0}'=0$ and $v _{\theta , 0}'=0$, then $R _0=1/2$ 
and $v _{ \theta,0}' =1/2$.
This is very similar to the Larson-Penston solution that describes
a spherical cloud having uniform density 
collapsing homologously, for which $R_0=2/3$
and $v _{\varpi,0}'=2/3$.

If $R _0 < 1/2$, small values of $b _{\theta ,0}'$ are not  allowed, 
whereas if $R _0 > 1/2$, small values of $v _{\theta ,0}'$ are forbidden.
In the first case, gravity is too weak and cannot produce the homologous 
collapse by itself, the condensation is induced by the toroidal pinching.
In the second case, gravity is too strong to produce the homologous collapse
and the cloud must be supported by the centrifugal force. 

These solutions are valid until the critical point is reached and describe 
a sub-Alfv\'enic (relatively to the similarity profile) dense core 
contracting homologously. 

\subsection{Critical point}
\label{crit_point}
A detailed analysis of the critical point can be achieved for the solutions
stated by Eqs.~(\ref{sol}). 
This has already been investigated for the Larson-Penston solutions by 
Whitworth \& Summers (1985) in the isothermal case and by 
Bouquet et al. (1985) for any values of $\gamma$. 

For the solutions stated by Eqs.~(\ref{sol}), we have:
\begin{eqnarray}
\nonumber
(v _\varpi - X)^2 -1 - b _\theta^2 / R - \Gamma _z ^2 R && = \\ 
( {1 \over 4} - (b _{\theta, 0}')^2 / R _0) X  ^2 && - 1 - \Gamma _z  ^2 R _0.
\label{critique}
\end{eqnarray}
Thus if $- \sqrt{R_0}/2 < b _{\theta,0}' < \sqrt{R_0}/2 $, the critical point occurs  at:
\begin{eqnarray}
X _c = -\sqrt{ { 1 + \Gamma _z ^2 R_0 \over 1/4 - (b _{\theta, 0}')^2/R_0  }  }, 
\label{crit_pos}
\end{eqnarray}
whereas if 
$(b _{\theta , 0} ')^2 > R _0/4$, there is no singular point since the 
 velocity relatively to the similarity profile is always smaller than the
Alfv\'en speed.

The  topological nature of the critical point can be studied by introducing
a new parameter, $s$,  such that:
\begin{eqnarray}
\nonumber
{d X \over ds } &=& D , \\
\nonumber
{d R \over ds } &=&  -R  N - { R \over X} D , \\
\label{param}
{d v _\varpi \over ds } &=&  (v _\varpi -X) N, \\
\nonumber
{d v _\theta \over ds } &=& - {v _\theta v _\varpi \over 
X (v _\varpi -X)} D, \\
\nonumber
{d b _\theta \over ds } &=&  -b_\theta N,
\end{eqnarray} 
where,
\begin{eqnarray}
\label{N_D}
N &=& 
\left(  {v _\theta ^2 \over X}  + { 1 \over X}    + R (v _\varpi -X) 
  + { \Gamma _z ^2 R  \over X} - {b _\theta ^2 \over X R}   \right) , \\
D &=& 
(v _\varpi - X)^2 -1 - b _\theta^2 / R - \Gamma _z ^2 R  .  
\nonumber
\end{eqnarray} 

The linearisation of the solutions stated by Eqs.~(\ref{sol}) near the
critical point, stated by Eq.~(\ref{crit_pos}) leads to a relation 
$d {\bf Y} / ds = [M] {\bf Y}$, where 
${\bf Y} =[X,R,v_\varpi, v_\theta, b _\theta]$ and 
where the matrix $[M]$ is given by the value of
 $\partial (d {\bf Y}_i / ds) / \partial  {\bf Y} _j$ at the critical point.
The matrix $[M]$ is easily calculated with Eqs.~(\ref{param}) and
 the derivative of $N$ and $D$.
We have:
\begin{eqnarray}
\partial D / \partial {\bf Y} &=& 
\left[X _c , {b _\theta^2 \over R^2} - \Gamma _z ^2  , - X _c , 0 , {-2 b ^c _\theta \over R_0} \right] , \\
\partial N / \partial {\bf Y} &=&
\left[ {-3 R_c \over 2} , {- X_c \over 2} +
 {\Gamma _z ^2 \over X_c} + {(b _\theta^c)^2 \over X_c R_c^2 } ,
 R_c  , {2 v _\theta ^c \over X_c} , {- 2 b _\theta ^c \over X_c R_c} \right]
\nonumber
\end{eqnarray}

%\begin{eqnarray*}
%[M] = \left(
%\begin{array}{ccccc}
%X _c & {b _\theta^2 \over R^2} - \Gamma _z ^2  & - X _c & 0 & {-2 b ^c _\theta \over R_0} \\
% R_c ({3 \over 2} R _c - 1) & R_c( {-2\Gamma _z ^2 R_c \over X_c} + {X_c \over
%2}) &  (-R_c + 1) R_c & {- 2 R_c
%v_\theta^c \over X _c} & {4 R _c b _\theta ^c \over X _c} \\
%3/4 R_c X_c & 1/4 X_c^2 - \Gamma _z ^2/2 - {(b _\theta^c)^2 / 2 R_c^2 } & {-R_c X _c \over 2} & - v _\theta ^c & b _\theta ^c \\ 
%v _\theta ^c   & - 2 {v _\theta ^ c \Gamma _z  ^2 R _c \over X _c} &
% - v _\theta ^c  
%& 0 & -2 {b _\theta ^c v _\theta ^c \over X_c} \\
%{ 3 \over 2} b _\theta ^c R _c  & b _\theta ^c ( {X_c \over 2}
% - 2 {\Gamma _z ^2 R _c \over X_c} ) &
%- b _\theta ^c R _c  & - 2 {b _\theta ^c v _\theta ^c \over X_c}  & 
%2 { (b _\theta^c)^2 \over X_c} 
%\end{array}
%\right) 
%\end{eqnarray*}
where $R _c$, $v _\varpi ^c$, $v _\theta ^c$ and $b _\theta ^c$ are the
values of the fields at the critical point. 

The study of the eigenvalues of $[M]$ allows to determine the topological
nature of the critical point.
The characteristic polyn\^omial $P$, of $[M]$ is:
\begin{eqnarray}
P(\lambda) = \lambda ^3 
\left(\lambda^2 - 
 X_c \lambda +  {X _c^2  \over 4} \left( 1 - 
{ 8 (b _{\theta,0}') ^4 \over  R _0^2 } 
\right) \right) =0. 
\label{poly}
\end{eqnarray}
Its roots are zero  (third order) and 
\begin{eqnarray}
\nonumber
\lambda _1 &=& {1 \over 2} X _c \left( 1 + 2 \sqrt{2} {(b _{\theta, 0}') ^2 \over R_0} \right), \\ 
\lambda _2 &=& {1 \over 2} X _c \left(1 - 2 \sqrt{2} { (b _{\theta, 0}') ^2 \over R _0} \right). 
\label{root}
\end{eqnarray}
0 is eigenvalue of $[M]$ at the third order since the vectors
 $(\partial (d {\bf Y}_i / ds) / \partial  {\bf Y})(X_c)$
 are linear combinations of $(\partial D / \partial {\bf Y})(X_c)$
and $(\partial N / \partial {\bf Y})(X_c)$ (see Eqs.~\ref{param}).

In the neighbourhood of the critical point, one has:
\begin{equation}
{\bf Y}(s) = \Sigma _{i=1,5} \alpha _i {\bf V} _i \exp(\lambda _i s)
\label{develop}
\end{equation}
where ${\bf V} _i$ are the eigenvectors and $\alpha _i$ are real numbers
and where $\lambda _3, \lambda _4$ and $\lambda _5$ are the three vanishing
eigenvalues.

Consequently, since  $\lambda _1$ and $\lambda _2$
 are both negative (remembering that $(b _{\theta, 0}')^2 / R _0< 1/4$),
 the critical points are nodes rather than saddles. The solutions
are thus able to cross the critical point.
It does not necessarily mean that the solutions are always physical, since
after this first critical point, it could be possible that the solutions
reach a second critical point through which it could be unable to pass.
Only a numerical study seems capable to give an answer to this question.

From Eqs.~(\ref{param}), we have the relations:
\begin{eqnarray}
\nonumber
dv _\theta &=& - { v _\theta ^c v _\varpi ^c \over (v _\varpi^c -X_c) X_c}
  dX , \\
\label{passage_crit}
d b _\theta &=& -  {b _\theta ^c \over v_\varpi^c -X_c} d v _\varpi, \\
d R &=& -  {R _c \over v _\varpi ^c - X_c} d v _\varpi - {R_c \over  X_c} dX,
\nonumber
\end{eqnarray}
where $v _\theta ^c = v _\theta(X_c)$, $v _\varpi ^c = v _\varpi (X _c)$ and
$R _c = R (X_c)$.

Thus, when the solutions cross the critical point, $v _\theta$, $b _\theta$
and $R$ vary according to these relations.

 When the solution crosses the critical point, 
$D$  becomes positive.
This leads to the condition:
\begin{eqnarray}
d V _\varpi > {X _c^2 +  R_0^2 \Gamma _z ^2 - (b_{\theta, 0}')^2 X _c ^2 / R _0
 \over X_c^2 + 2 R _0^2 \Gamma _z ^2 - 
 (b_{\theta, 0}')^2 X _c ^2 / R _0 } d X.
\label{condi}
\end{eqnarray}

According to Eqs.~(\ref{passage_crit}) and~(\ref{condi}), and as  
pointed out by Whitworth \& Summers (1985), there is a one dimensional
space of possible solutions for $X< X _c $ which are compatible with a
given solution for $X > X_c$.

\subsection{Expected behaviors}
\label{blabla_crit}
The homologous solutions stated by Eqs.~(\ref{hypo}) and~(\ref{sol}) 
allow us to anticipate some trends that we will 
 recover numerically in the next section. 

 From Eq.~(\ref{crit_pos}), it is seen that, $|X _c|$ increases with 
$\Gamma _z  R _0 = b _z$ and $b _{\theta, 0}'$.  From Eqs.~(\ref{hypo}), it is 
seen that the value of $v _\varpi(X_c)$, $v _\theta(X_c)$ and
 $b _\theta(X_c)$ increases
with $|X _c|$, whereas $R(X_c)$ and $b _z(X_c)$ do not depend on $|X _c|$.

Consequently, the following trends are expected:
\begin{description}
\item[-] If the density  increases, $|X_c|$, 
the position of the magnetosonic point,
 decreases (remembering that $b _z=\Gamma _z  R$). 
This is simply due to the fact that the Alfv\'en speed decreases and
thus the velocity of the fastest waves decreases as well. 
With Eqs.~(\ref{sol}),
it is seen that the rotation ($v _{\theta, 0}'$) increases.
This means that when density increases, the gravitational force
increases as well and must be counterbalanced by the centrifugal force.
\item[-] If the poloidal magnetic intensity increases,
 $|X _c|$
increases and the value of the radial velocity
and toroidal magnetic field at this point increase as well.  

Consequently, the trend is that the larger the magnetic poloidal 
intensity is, the larger $| v _\varpi |$, $| v _\theta |$ and $ |b _\theta |$
  will be in the super-Alfv\'enic
part ($X < X _C$). 
This is simply due to the fact that the magnetic poloidal pressure 
 leads to an effective sound speed equal to: $1 + \Gamma _z ^2 R$
(see Eq.~\ref{eqfin1}) and that $V _\varpi, V _\theta$ and  $B _\theta$
are proportional to the sound speed.

\item[-] If the toroidal field increases, the position of the critical
point increases and the value of the radial velocity as well, 
which simply means that the cloud is more compressed by the toroidal
pinching. 
As we already said, if $(b _{\theta, 0}')^{1/2} > R_0/2 $ there is no 
critical point since the velocity of the fastest waves is always
greater than the velocity of the fluid relatively to the similarity profile.
Rotation must increase (Eq.~\ref{sol}) in order to counterbalance the
 toroidal pinching.
\end{description}

An important parameter is the value of the radial velocity at large 
$X$. According to Eq.~(\ref{asymptote}), the radial velocity slowly
diverges when $X \rightarrow -\infty$ but, as we already said, the
physical  meaning of the solutions is unclear in this limit. 
On the other hand, Eqs.~(\ref{cond}) and~(\ref{crit_pos}) give for 
$v _\varpi(X_c)$ the value $X_c /2 \le -1$ (in sound speed unit).
 Consequently, it is
expected that the typical radial velocity at few $X _c$ should be 
 as high as few times the value of the sound speed.

The smallest value of $|v _\varpi (X _c)|$ is 1 and is reached 
for the unmagnetized solution. In that case $X _c=-2$.
Let us recall that for the Larson-Penston solution, $X _c=-3$
and $v _r(X _c)=-2$. The present unmagnetized solutions are thus less 
dynamical, in the inner homologous part and in the radial direction,
 than the Larson-Penston solution.

\section{Numerical results}

\subsection{Method}
In this section we investigate numerically the system of 
Eqs.~(\ref{eqfin1})-(\ref{eqfin4}). These equations 
are four ordinary differential
equations that can be easily integrated by a fourth order 
Runge-Kutta method.

At $X=0$, the system is singular and one has to develop the solutions
 to first order and to start the integration at $X \ne 0$.
The asymptotic development towards $X \rightarrow 0$ is given by 
Eqs.~(\ref{hypo}) and $v _ {\varpi, 0}' = 1/2$  (Eq.~\ref{cond}).
Consequently, 
 the  number of free parameters
is four ($R _0$, $v _{\theta, 0}'$, $\Gamma _z $ and $b _{\theta, 0}'$) 
and there is
one more free parameter when the solutions cross the critical point, 
leading to a total of five free parameters. 
It is thus difficult to explore all the parameter space and we will restrict
 the investigation to a limited parameter range.
We start at $X = -10^{-5}$ and use a spatial step $dX=-10^{-3}$.

The integration through the critical point requires some care. 
As we already said, the solutions for $X < X _c$ compatible with 
a solution for $X > X_c$ are not unique and are fixed by 
Eqs.~(\ref{passage_crit}) and by the conditions stated by Eq.~(\ref{condi}).
We proceed as follow. When  
 the critical point is reached ($D$ reverse sign), we select a value of 
$d v _\varpi$ satisfying Eq.~(\ref{condi}) and give to the other fields
the values stated by Eqs.~(\ref{passage_crit}).  
In the following we  restrict our study to  $d v _{\varpi} =0$. 

We will first consider homologous cores (thus following Eq.~\ref{sol})
and then relax this assumption and consider the more general case of 
non-homologous cores.

%As it has been pointed out
%by Whitworth \& Summers (1985) the outward integration is sometime unstable 
%and we are restricted most of the time to $d v_\varpi > v _\varpi / X_c dX$.
%Although in  principle, it could be  possible to proceed as 
%Whitworth \& Summers who restart integration
% from $X < X _c$ and integrate inward in 
%order to find all solutions that reach the critical point,  
%in the present study we have four variables which leads to a larger
%set of possible parameters and makes  this approach difficult. 

%We find that varying $d v _\varpi$ in the stable numerical integration 
%domain has little influence on the solutions except if $d v _\varpi$
%is chosen large, i.e. greater than $\simeq$10-100. However, as
%it has been pointed out by Whitworth \& Summers (1985), 
%this kind of solutions involving shocks  are forbidden by the second law
%of thermodynamics, since they entail gas flowing into the discontinuity 
%subsonically and out supersonically (these solutions are allowed 
%for the expansion solutions).
%For all these reasons, we restrict our study to  $d v _{\varpi} =0$. 

\subsection{Homologous cores and non homologous super-Alfv\'enic envelope}

\begin{figure}
\includegraphics[width=8cm]{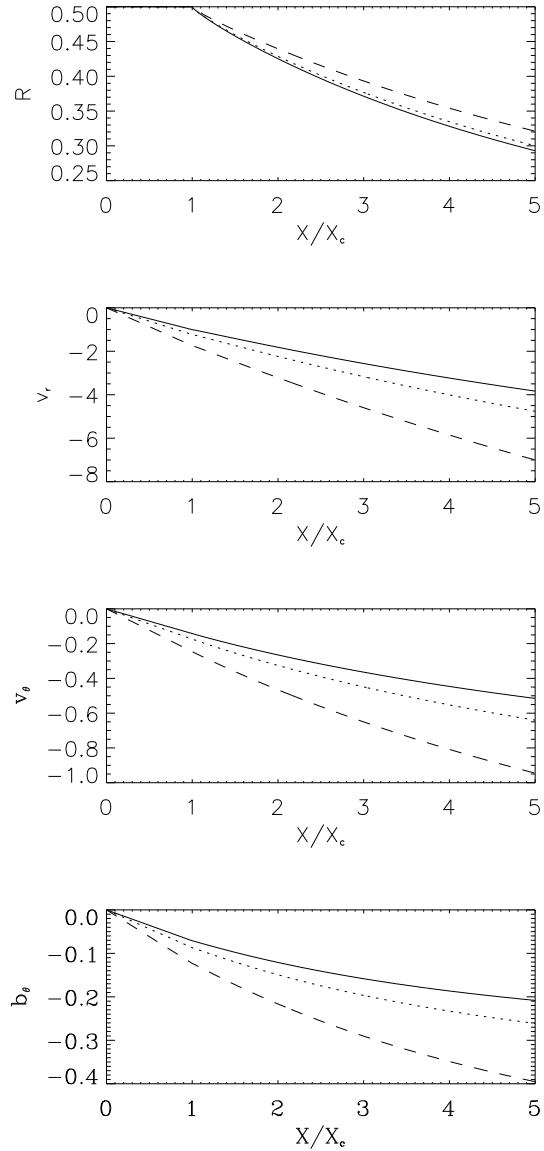}
\caption{Solutions corresponding to $R_0=1/2$, $v _{\varpi,0}'=1/2$,
$b _z= 0$ (full line) $0.5$ (dotted line) and $1$ (dashed line) 
for $b _{\theta,0}'=0.1 \sqrt{R _0/4}$. The bifurcation occurs at $X=X_c$
according to Eq.~(\ref{crit_pos}). }
\label{bthet0.1}
\end{figure}

\begin{figure}
\includegraphics[width=8cm]{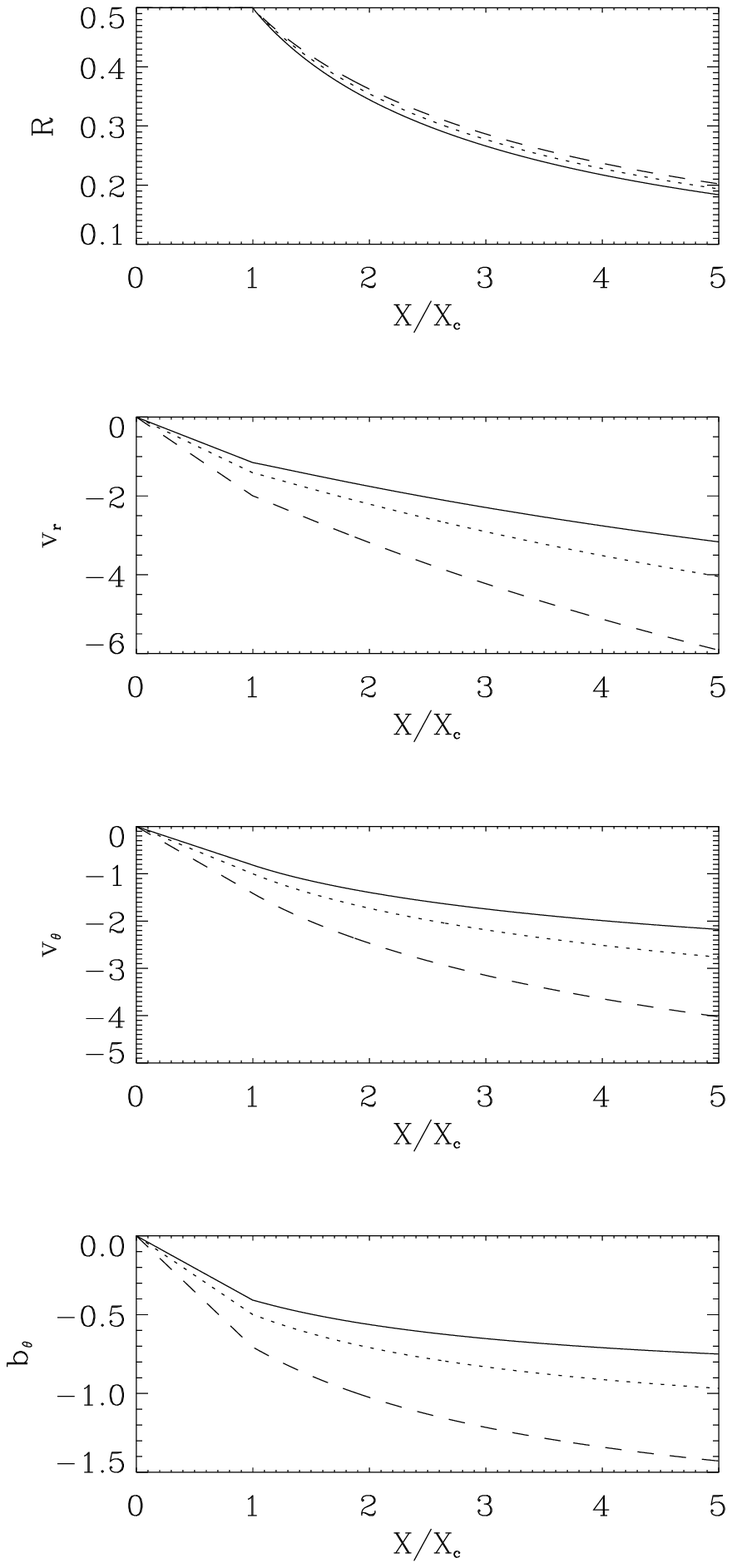}
\caption{Same as Fig.~\ref{bthet0.1} for $b _{\theta,0}'=0.5 \sqrt{R _0/4}$.}
\label{bthet0.3}
\end{figure}

\begin{figure}
\includegraphics[width=8cm]{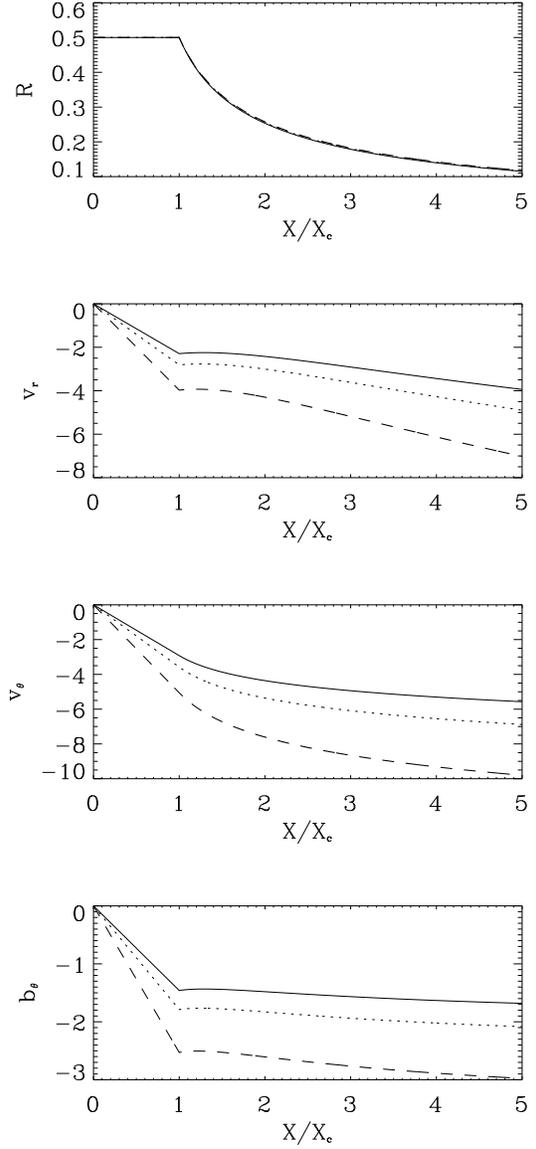}
\caption{Same as Fig.~\ref{bthet0.1} for $b _{\theta,0}'=0.9 \sqrt{R _0/4}$.}
\label{bthet0.45}
\end{figure}

The homologous solutions stated by Eqs.~(\ref{hypo}) and~(\ref{sol}) play
an  important r\^ole since except for the geometry 
they are equivalent to the Larson-Penston 
solution that has been found to be in reasonable agreement with numerical 
simulations in previous studies (Larson 1969, Hunter 1977, 
Blottiau et al. 1988, Foster \& Chevalier 1993). 
Moreover, these solutions are explicit and allow us
to check  carefully  the validity of the numerical integration.
 
In this section we present  the numerical integration of 
Eqs.~(\ref{eqfin1})-(\ref{eqfin4}) with the initial density $R _0 = 1/2$. 
This value is important since according to Eq.~(\ref{sol}),  
gravity is exactly equal to the advection term and all values of
$v _{\theta, 0} '$  and $b _{\theta, 0} '$ are allowed.

The other parameters are:
$v _{\varpi,0}'=1/2$ and in order to explore a large range of
parameters $b _z=0, 0.5$ and $1$ corresponding to $\Gamma _z =0,\, 1$ and $2$
respectively and   $b _{\theta,0}'$ equal to 
$\alpha \sqrt{R_0/4}$, whith $\alpha=0.1, 0.5$ and 0.9. 
$v _{\theta,0}'$ is calculated according to Eqs.~(\ref{sol}).
We integrate until $X  = 5 X _c$ and display the solutions as a function 
of $X / X _c$ (see Eq.~\ref{crit_pos}). Thus, the bifurcation always
occurs at $X / X _c=1$.

Before presenting the numerical results, we briefly 
discuss the  field intensities.
\subsubsection{Field intensity}
For orientation, we estimate the typical values of 
the poloidal field in the subsonic homologous core and
the value of the toroidal component at the critical point
as a function of $\alpha$.
With Eq.~(\ref{champ})-(\ref{crit_pos}), one has:
\begin{eqnarray}
B _z = \sqrt{\mu _0 C _s ^2 \rho} \Gamma _z \sqrt{R _0},
\end{eqnarray}
\begin{eqnarray}
B _\theta = \sqrt{\mu _0 C _s ^2 \rho} 
\sqrt{ {\alpha^2 \over 1 - \alpha ^2} }
 \sqrt{1 + \Gamma _z^2 R _0}. 
\end{eqnarray}
If a canonical density of $10^6$ cm$^{-3}$
 in the collapsing filament is adopted and for
$R _0 = 1/2$, $\Gamma _z=1$, then typically in the weak,
intermediate and strong toroidal field cases, toroidal 
field strengths of $\simeq$15, $\simeq$90 and $\simeq$320 $\mu G$ 
are associated with 
poloidal field of about $\simeq 90 \mu G$.

\subsubsection{Weak toroidal magnetic  field}
Fig.~\ref{bthet0.1} displays the results for $b _{\theta,0}'=0.1
\sqrt{R _0/4}$.

For $\Gamma _z=0$, the solution is weakly magnetized 
(since the toroidal field is weak) and except for the geometry it is equivalent
to the Larson-Penston solution.  As pointed out by Whitworth \& Summers (1985)
this solution describes a strong compression wave propagating inwards 
into the cloud, the bifurcation point being the head of this compression
 wave. 

It is seen that for $X/X_c < 1$ the solution is homologous as predicted
by Eqs.~(\ref{sol}). At $X/X_c>1$ the density
decreases, whereas $|v _\varpi|$, $|v _\theta|$ and $|b _\theta|$ increase.
The density decreases less rapidly for large values of $b _z$ 
($\Gamma _z $) and
as discussed in Sect.~\ref{blabla_crit}, the velocity and the toroidal
magnetic field increase with $b _z$ (see Eq.~\ref{hypo} and~\ref{crit_pos}).
 
The solutions are weakly peaked with $R(0)/R(5 X_c) \le 2$. The velocity
and toroidal magnetic fields continue to decrease rapidly
  after the critical point. We have 
$v _\varpi(X_c) / v _\varpi(5 X_c) \simeq 0.25$.

The radial velocity is roughly $8$ times larger than the azimuthal one. 
 When $b_z=0.5$ and $1$  the poloidal magnetic intensity dominates
the toroidal magnetic intensity 
in the core (by a factor $\simeq 5$ at $X_c$ for $\Gamma _z =1$,
 and $\simeq 7$ for $\Gamma _z =2$). At $X \simeq 5 X_c$, 
the two components are almost equal for $\Gamma _z =1$ whereas for 
$\Gamma _z =2$, $b _z / b _\theta \simeq 1.5$.

The radial velocity at $5 X_c$ is around $4-5 \, C_s$ for low and intermediate
poloidal magnetic intensity ($\Gamma _z =0$ and 1) and
 it is much higher ($\simeq 7$) for a strong poloidal magnetic
field ($\Gamma _z =2$).

\subsubsection{Intermediate toroidal magnetic  field}
The case $b _{\theta,0}'=0.5 \sqrt{R _0/4}$ is displayed in 
Fig.~\ref{bthet0.3}.
The toroidal magnetic force in the core ($2 b _{\theta,0}' X / R_0$)
 is about half of the gravitational force ($R _0 X /2$). Thus the toroidal
field is important but not dominant.

It is seen that the bifurcation at the magnetosonic point is stiffer. 
The density decreases more rapidly ($R(0)/R(5 X_c) \le 2.5$) than in the 
previous case. This is due to the fact that the cloud is compressed 
by the toroidal magnetic field.

The velocity and the magnetic field evolution is much 
 flatter in the envelope than in the core,
we have $v _\varpi(X_c) / v _\varpi(5 X_c) \simeq 0.3$.
Thus $v _\varpi(5 X _c)$ is even slightly lower than for the case with
weak toroidal magnetic field in spite of the higher value of $v _\varpi(X_c)$.

 It is also seen that
the density variation with the poloidal magnetic intensity is small
compared to the variation of the velocity and the toroidal magnetic 
component ; we discuss this further in the next section. 

The amplitude of the radial velocity is about $2/3$ the amplitude of
the azimuthal one.
 The poloidal magnetic component dominates 
the toroidal one in the core (ratio 0.5-0.75 at $X=X_c$). 
The two component become comparable at $X \simeq 1.5 X_c$ which is
 earlier than for the weak toroidal magnetic field. At $X / X_c=5$
$b _z / b _\theta \simeq 3$.

\subsubsection{Strong toroidal magnetic  field}
The case $b _{\theta,0}'=0.9 \sqrt{R _0/4}$ is displayed in 
Fig.~\ref{bthet0.45}.
The toroidal force is larger than the gravitational force 
(factor $\simeq 1.62$) and the solution describes the inward propagation
of a strong magnetic compression wave into the cloud.

The most striking features of the solutions with strong toroidal magnetic field
  is that the density is almost 
independent on the poloidal magnetic intensity, whereas the other fields
are roughly proportional to $\Gamma _z $. This is due to the fact, that since the
toroidal field is strong ($b _\theta^2 (X_c) / R   > 3$), the magnetic and 
centrifugal forces are large compared to the thermal pressure. Thus 
in Eq.~(\ref{eqfin1}), the thermal pressure can be neglected 
(term $-1$ in $D$ and $1/X$ in $N$). In this limit, 
it is then seen that Eq.~(\ref{eqfin1})
is invariant under the rescaling: 
\begin{eqnarray}
\nonumber
R &\rightarrow& R, \\
\nonumber
v _\varpi &\rightarrow& K  v _\varpi, \\
\label{rescale}
v _\theta &\rightarrow& K  v _\theta, \\
\nonumber
b _z &\rightarrow& K  b _z, \\
\nonumber
b _\theta &\rightarrow& K  b _\theta.
\end{eqnarray}

Another interesting feature is that the radial velocity and the toroidal 
 magnetic field are almost constant just after the critical point. They then
decrease significantly again around $X / X_c \simeq 1.5$. 
The radial velocity then 
decreases according to Eqs.~(\ref{asymptote}) whereas $v_\theta$ and
 $b_\theta$ decrease more slowly and rapidly 
 converge toward a constant value.

The solutions are even more peaked than in the previous cases, with
$R(0) / R(5 X_c) \simeq 5$ whereas the radial velocity is  flatter with
$v _\varpi( X_c) / v _\varpi(5 X_c) \simeq 0.5$.

The azimuthal velocity is  about $1.3-1.5$ larger than the radial velocity. 
The  toroidal magnetic field rapidly dominates the poloidal one since 
at $X/X_c \simeq 0.5$, the two field intensities are comparable. At 
$X/X_c$, $b _\theta/ b_z \simeq 2$ and at $X/X_c =5$, we have
$b_\theta / b_z \simeq 10$.

For $\Gamma _z =0$ or $\Gamma _z =1$, $v _\varpi(5 X_c) \simeq 4-5 \, C _s$.

\subsubsection{Varying the density: forbidden  values}
\label{interdit}
\begin{figure}
\includegraphics[width=8cm]{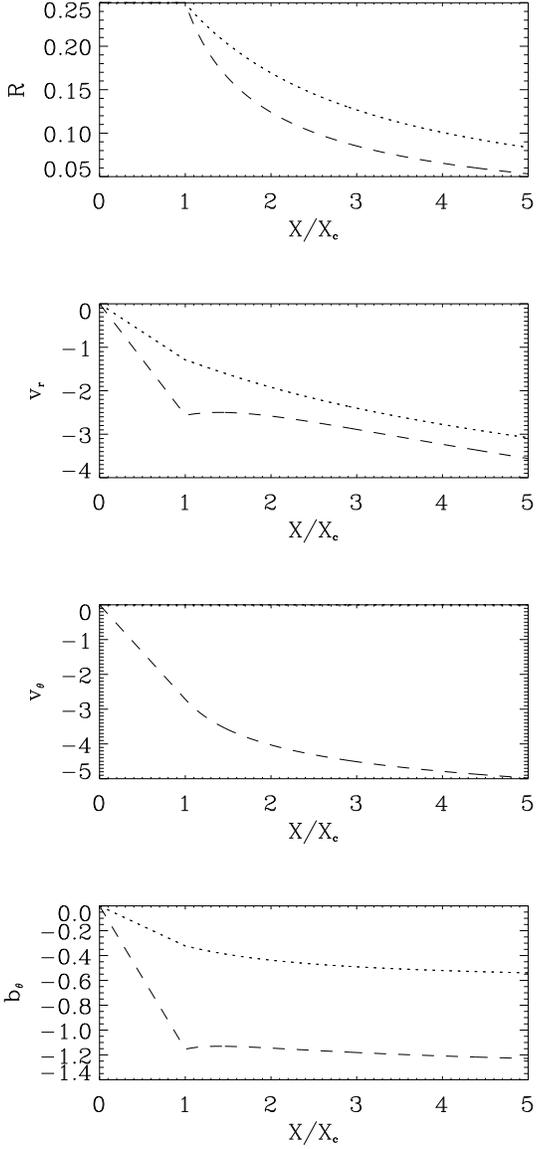}
\caption{Solutions corresponding to $R_0=0.25$ and $b_{z,0}=0.25$
($\Gamma _z =1$) for $b_{\theta,0}'=0.5 \sqrt{R _0/4}$ (dotted line) and
 $b_{\theta,0}'=0.9 \sqrt{R _0/4}$ (dashed line).}
\label{R0.25}
\end{figure}

\begin{figure}
\includegraphics[width=8cm]{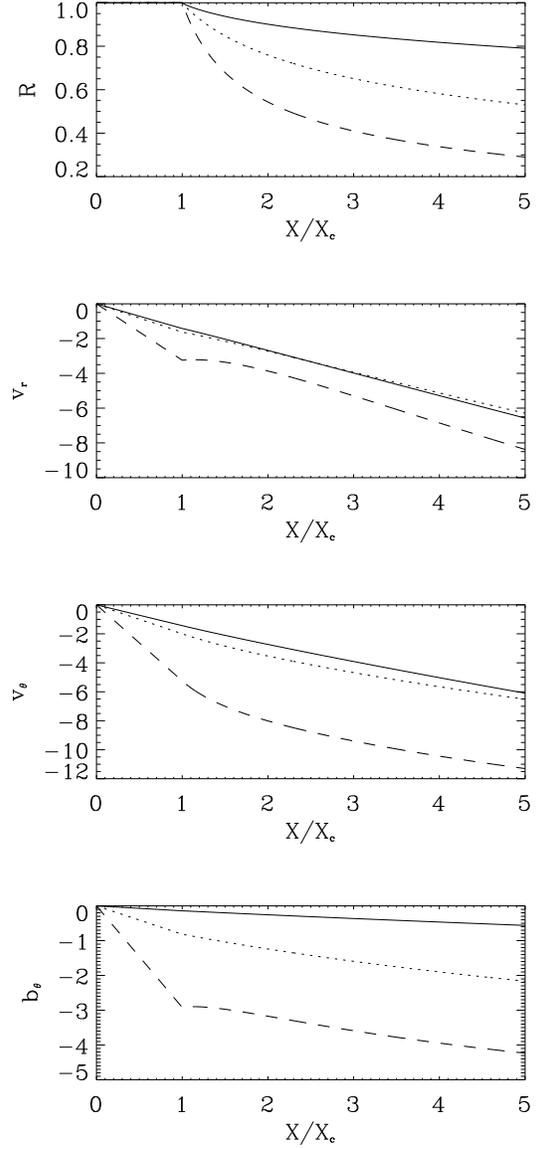}
\caption{Solutions corresponding to $R_0=1$ and $b_{z,0}=1$
($\Gamma _z =1$) for $b_{\theta,0}'=0.1 \sqrt{R _0/4}$ (full line),
 $b _{\theta,0}'=0.5 \sqrt{R _0/4}$ (dotted line) and 
$b  _{\theta,0}'=0.9 \sqrt{R _0/4}$  (dashed line).}
\label{R1}
\end{figure}

In this section, we vary the density of the homologous core in order 
to explore the dependence of the fields on this parameter.
We look carefully at two particular cases and briefly comment on other
values. 
In order to make comparison with previous cases easier we choose
$\Gamma _z =1$ (dotted line of previous section) and $b _\theta=0.1 \sqrt{R
_0/4}, \, 0.5 \sqrt{R _0 /4 }$ and $0.9 \sqrt{R _0/4}$, i.e weak, intermediate
and strong magnetic toroidal intensity.

The two cases, $R_0=1/4$ and $R_0=1$
 are more closely considered. 
In the first case, gravity alone cannot explain the collapse 
(see Eq.~\ref{sol}). The collapse is thus assisted by the toroidal
magnetic field. In the second case, gravity is too strong and must be
partially  counterbalanced by the centrifugal force.

The results are 
respectively displayed in Fig.~\ref{R0.25}
and Fig.~\ref{R1}. The full line is $b_\theta=0.1  \sqrt{R
_0/4}$ (this case is not
possible for $R_0=1/4$ according to Eqs.~\ref{sol}), the dotted line
$b_\theta=0.5 \sqrt{R_0 / 4}$ and the dashed line $b _\theta=0.9
\sqrt{R _0 / 4}$.

From a comparison between Fig.~\ref{R0.25} and~\ref{R1}, it is seen 
that the  density is more peaked whereas the  radial velocity
is flatter for small densities ($R_0=1/4$) than for large one ($R_0=1$). 

For $R _0=1/4$, the azimuthal velocity vanishes according to Eq.~(\ref{sol})
 when $b _\theta=0.5 \sqrt{R _0/4}$.
When, $b _\theta=0.9 \sqrt{R _0/4}$, it is seen that the radial velocity
and the toroidal field
increase slightly just after the critical point,
 then reduce  at $X/X_c \simeq 1.5$. \\

Smaller values of density ($R _0 < 1/4$), 
require values of $b _{\theta,0}'$ 
 closer to $\sqrt{R _0/4}$
 in order to have positive values of $| v _\theta |$
(see Eqs.~\ref{sol}). Similar behaviors to the previously presented
cases are obtained.

For larger values of density ($R _0 > 1$), we find that with
 $b _{\theta,0}'=0$, 
the solutions oscillate around the homologous solution, i.e. after the 
critical point the density decreases and increases alternately. When 
$b _{\theta,0}' \ne 0$ (e.g. $b _{\theta,0}' > 0.1$)
 then the solutions cross
another critical point and we stop the numerical integration. 
 This behavior is somehow similar to the results obtained by 
Whitworth \& Summers (1985) who find that the allowed values of the density
at the origin in the Larson-Penston equations are quantized. 
It could thus be possible that larger values of $R_0$ are allowed in 
Eqs.~(\ref{eqfin1})-(\ref{eqfin4}). Only a systematic numerical search
could answer this question.

\subsection{Non-homologous cores}

\begin{figure}
\includegraphics[width=8cm]{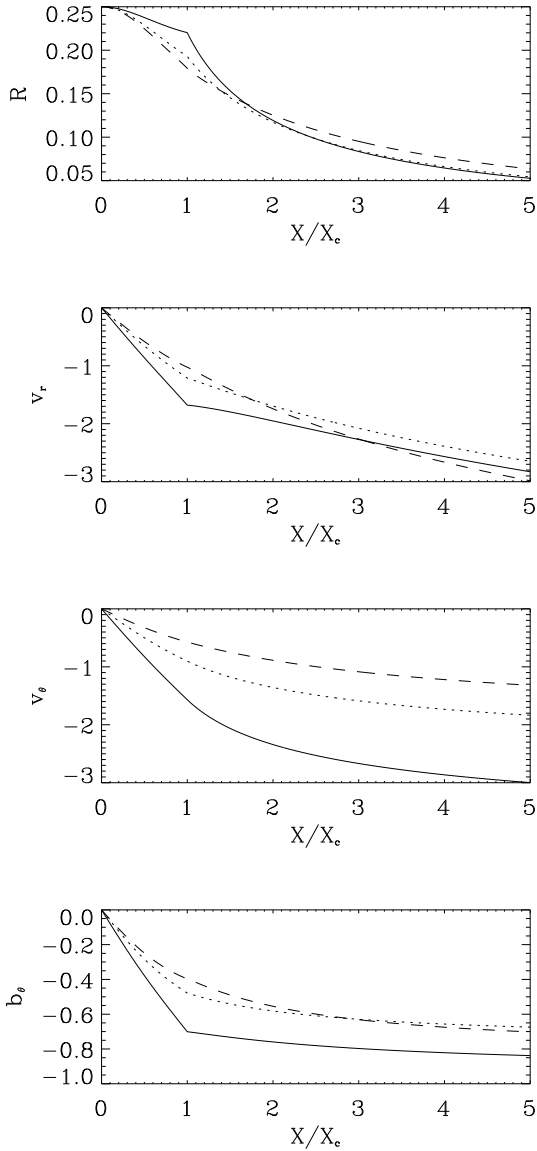}
\caption{Solutions corresponding to $R _0=0.25$, $b _{z, 0}=0.25$,
$b _{\theta,0} ' =0.9 \sqrt{R _0/4}$. Full line is $\delta v _\theta'=-0.05$, 
 dotted line is $\delta v _\theta'=-0.15$ and 
 dashed line  $\delta v _\theta'=-0.25$ (see text).}
\label{nhR0.25}
\end{figure}

\begin{figure}
\includegraphics[width=8cm]{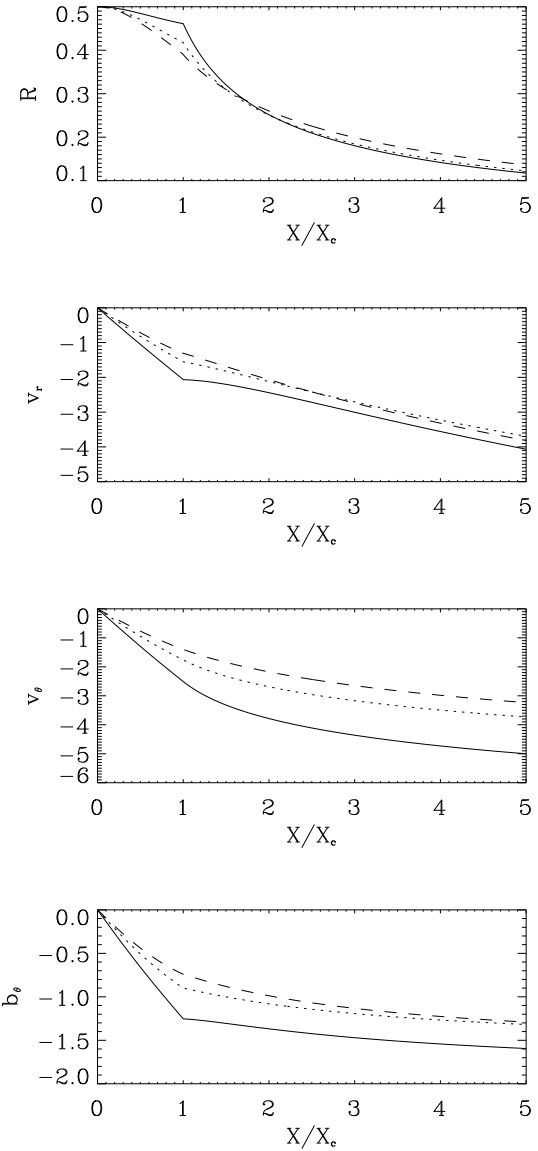}
\caption{Solutions corresponding to $R _0=0.5$, $b _{z, 0}=0.5$,
$b _{\theta,0}'=0.9 \sqrt{R _0/4}$. 
Full line is $\delta v _\theta'=-0.02$, 
 dotted line is $\delta v _\theta'=-0.06$ and 
 dashed line  $\delta v _\theta'=-0.1$ (see text).}
\label{nhR0.5}
\end{figure}

In this section, we investigate the case of non homologous cores, i.e.
we start with initial conditions that do not satisfy the second 
 conditions stated by Eqs.~(\ref{sol}).
Let us introduce the parameter, 
$\delta v_\theta '= v _\theta ' - v_{\theta,0}'$, where $v _{\theta,0}'$ is
the value stated by Eqs.~(\ref{sol}). The smaller $|\delta v _\theta '|$ 
is, the closer we are to the homologous solutions.

We consider the two cases $R _0=0.25$ ; and $R _0=0.5$ with $\Gamma _z =1$ and 
$b _{\theta,0}'=0.9 \sqrt{R_0/4}$ for three values of $\delta v _{\theta}'$ 
($-0.05, \, -0.15, \, -0.25$ for $R _0=0.25$ and $-0.02, \, -0.06, \, -0.12$
for $R _0 = 0.5$). The results are displayed in Figs.~\ref{nhR0.25}
and~\ref{nhR0.5}. Full lines represent the smallest value of 
$| \delta v _\theta ' |$,
dotted line, the intermediate value, and dashed line the largest.

It is seen that the cores are no longer homologous and become
 less flat when $|\delta v _\theta '|$ increases. Also, the transition 
between the sub-Alfv\'enic part and the super-Alfv\'enic one, is smoother.
The poloidal force does not vanish any more in the cores and acts
explicitly  everywhere.

For $R_0=0.25$,
the density and radial velocity values at $5 X _c$ do not vary
 significantly with $\delta v_\theta'$, whereas the values of 
the azimuthal velocity and the toroidal magnetic field decrease
significantly  when $|\delta v _\theta '|$ increases.

For larger values of $|\delta v _\varpi'|$ 
 the solutions (for $R _0=0.25$ and $R _0=0.5$) 
cross another critical point. 
We consider them as not physical and we stop the numerical integration.
It is then not possible for these two cases
to obtain solutions with a vanishing azimuthal velocity as 
suggested by Eq.~(\ref{sol}).

For larger values of $R _0$, the largest allowed value of 
$\delta v _\varpi'$ decreases rapidly (for $R _0=1$, we find that the 
largest allowed $|\delta v _\theta '|$ is between $0.03$ and $0.04$).
This is consistent with the fact that the homologous solutions with
large  $R _0$ are not physical (see Sect.~\ref{interdit}).

It is worth noting that the value of the radial velocity at $5 X _c$
is around 2.5 to 3 which is lower than the values obtained
previously. The smallest value of $v _\varpi (5 X _c)$ we found is
around 2 and is obtained for $R _0 < 1/8$, $b _\theta \simeq
\sqrt{R _0/4}$, $\Gamma _z  < 1$ and $|v _{\theta,0}'| < 0.2$. 

This value
is still  3 to 4 times higher than the value observed by Tafalla et
al. (1999) in the starless cores L1544.
However, it is known from previous study (e.g Foster \& Chevalier 1993)
that the Larson-Penston solution 
(in which the  velocity tends to $3.3 C _s$ at large $r$)
is in good agreement with the inner collapsing
part of the cloud, the external part being significantly different. 
This is due to the fact that the self-similar solutions make the assumption
that the system is infinite and is self-similar everywhere. In particular, 
the self-similar assumption does not take into account the external medium. 
Self-similar solutions thus appear in the inner part of the cloud when 
the boundary conditions have been {\it forgotten} by the system.

\section{Conclusion}
In this paper, 
we  have derived self-similar solutions able to describe the inner part
of a collapsing,  rotating, 
magnetized, self-gravitating, isothermal filaments. 
The filaments is collapsing  homologously  along the  z-axis, 
the slope of the  axial velocity being two times the slope of the 
radial one at the origin.   
 The four ordinary equations obtained are similar to the 
equations derived by Penston (1969) and Larson (1969). They 
present a critical magnetosonic point that induces a bifurcation.

For some of the homologous solutions already obtained previously, 
(Aburihan et al. 2001, Hennebelle 2001)
we carried out a study of the critical point and found that there is 
a critical value of $b _{\theta, 0}'$ 
(slope of the toroidal field at the origin) 
above which no bifurcation occurs. 
By studying the eigenvalues
of the linearized equations  in the neighbourhood of the critical point,
 we demonstrate that these solutions are able to cross the singularity since
it is a node rather than a saddle point. 

We have then explored the 
system numerically and obtained a series of density, velocity and
magnetic fields 
profiles that could help to understand the observational data and could
be used as benchmarks for a full numerical simulation.
% The following trends are obtained~:
%\begin{description}
%\item[-] In the central core  the poloidal magnetic field is higher than the 
%toroidal one whereas in   the envelope, the toroidal field dominates.
%\item[-] The stronger the toroidal field
% (but not above the critical value), the more peaked is the density
%and the flatter are the azimuthal velocity and toroidal magnetic field in the 
%envelope.
%\item[-] The values of the radial and azimuthal velocities and of the toroidal 
%magnetic field increases with $\Gamma _z $ (where $b _z=\Gamma _z  R$) for a given value of 
%$R(0)$ (the central density) and with $R(0)$ for a given value of $\Gamma _z $. 
%\end{description}
% We found also that some solutions are not physical as
% for large initial densities the solutions oscillate around the homologous
%solutions and sometimes reach another critical point through which they
%are unable to pass.  

The  solutions obtained in this paper have some restrictions. 
First, they are valid
not too far from the equatorial plane since they have an homologous 
axial velocity field that quickly diverges and not too far 
(few $|X_c|$) from the z-axis since the axial component of the
velocity field  does not depend on $\varpi$.

%The solutions predict a radial velocity which is too high (factor 3-4)
% and cannot
%be compared to the present observations (Tafalla et al. 1999). This
%could be due to the above mentioned restrictions or to the
%self-similar assumption itself.

 Nevertheless, these solutions are the first 
semi-analytical solutions  describing the condensation of a
magnetized rotating filament and  having complex (non homologous)
spatial profiles.
They can be used for future analytical or
numerical studies of the gravo-magnetic condensation.

\section{acknowledgment}
I thank Anthony Whitworth and Derek Ward-Thompson for stimulating discussions.
I thank D. Moss, the referee, for his help in improving the manuscript.  
I gratefully acknowledge the support of an European Commission 
Research Training Network under the Fifth Framework Programme (No. 
HPRN-CT2000-00155).

\end{document}